\documentclass[%
 reprint,
superscriptaddress,nofootinbib,
 amsmath,amssymb,
 aps]{revtex4-2}

\usepackage{lmodern}

\usepackage[colorlinks=true,linkcolor=red,citecolor=blue,urlcolor=blue]{hyperref}

\usepackage{graphicx} 
\usepackage{amsmath,amssymb,physics}
\usepackage{cleveref}
\usepackage{orcidlink}

\usepackage{xcolor}

\begin{document}

\preprint{APS/123-QED}

\title{Distinct Signatures of the Nature of Phase Transition in Binary Neutron Star Mergers}

\author{Sagnik Chatterjee \orcidlink{0000-0001-6367-7017}}
\email[Contributed equally: ]{sagnik.chatterjee@nipne.ro}
\affiliation{Indian Institute of Science Education and Research Bhopal, Bhopal 462066, India}
\affiliation{National Institute for Physics and Nuclear Engineering (IFIN-HH), RO-077125 Bucharest, Romania}

\author{Shamim Haque \orcidlink{0000-0001-9335-5713}}
\email[Contributed equally: ]{shamims@iiserb.ac.in}
\affiliation{Indian Institute of Science Education and Research Bhopal, Bhopal 462066, India}

\author{Kamal Krishna Nath \orcidlink{0000-0002-4657-8794}}
\affiliation{School of Physical Sciences, National Institute of Science Education and Research, An OCC of Homi Bhabha National Institute, Jatni-752050, India}

\author{Rana Nandi \orcidlink{0000-0002-6277-2618}}
\affiliation{Department of Physics, Indian Institute of Technology Delhi, New Delhi 110016, India}

\author{Ritam Mallick \orcidlink{0000-0003-2943-6388}}
\email{mallick@iiserb.ac.in}
\affiliation{Indian Institute of Science Education and Research Bhopal, Bhopal 462066, India}

\date{\today}

\begin{abstract}
Binary neutron-star mergers offer crucial insights into the matter properties of neutron stars. We present the possible imprints in the gravitational wave signal from the nature of phase transition from such events. Our study employs a one-parameter family of equation of states built using a polytropic approach with a control parameter $\Delta p$ surveying the features of hadron-quark phase transition, from Maxwell construction to the Gibbs construction. It allows us to explore the extent of mixed phases and analyse their direct impact on merger dynamics. Post-merger gravitational wave emissions reveal the expression of specific signatures in the spectrogram and power spectral density, serving as a distinct signature of equations of state with mixed phases. We found additional peaks in power spectral density that are exclusively generated from the post-merger remnant experiencing a phase transition. Additionally, the nature of phase transition leaves specific imprints on the spectrogram, leading to a two-folded signature from gravitational wave analysis. Furthermore, we establish the first correlation between $\Delta p$ and the threshold mass for prompt collapse. Our analysis shows that $\Delta p \lesssim 0.04$ is required if GW170817 formed a long-lived remnant or has experienced a delayed collapse into a black hole.
\end{abstract}

\maketitle

\section{Introduction}\label{sec:intro}

Calculations from Quantum Chromodynamics (QCD) predicts the existence of gluons and quarks in deconfined state at asymptotically high densities \cite{Fraga_2014,Fadafan_2020} and hints towards a possible phase transition (PT) \cite{Dexheimer,Somasundaram} from hadronic matter (HM) \cite{Tolos_2024, Providencia_2024} to quark matter (QM) \cite{Weber_1999, Annala_2020}. These density regimes are yet to be probed by terrestrial observations, making Neutron Stars (NSs) paradigmatic for studying matter at such densities. NSs are dense compact objects found in the universe having central densities reaching up to 2--8 times that of the nuclear saturation density ($n_0 \approx 0.16~\mathrm{fm}^{-3}$) \cite{glendenning2012compact}. The mass and radius measurement of a few NSs in the last decade has remarkably improved our understanding of the equation of state (EoS) at high-density, low-temperature regimes~\cite{Antoniodis_2013,Fonseca_2021,Cromartie_2020,Miller_2019,Riley_2019,Miller_2021,Riley_2021}. Living inside these observational constraints, a significant amount of literature has attempted to understand the feasibility of the PT process in isolated NSs through equilibrium models~\cite{Glendenning_1992,Alford_2001,Alford_2005,Benic_2015,olinto_1987,Bhattacharyya_2006,Bhattacharyya_2007, Drago_2007,Baldo_2003,Blaschke_2009,Dexheimer_2010,Orsaria_2014,Ferreira_2020,Han_2020, Tewari:2024qit,Debanjan_2024,Thakur:2024ijp} and dynamical evolutions~\cite{Dimmelmeier_2009,Franzon_2016,Prasad_2018,Prasad_2020,Liebling_2021,Shashank_2023,Espino_2022,Naseri_2024}. The merging events of these compact objects supply additional information about matter at its core. The first-ever detection of gravitational waves (GWs) in GW170817 \cite{Abbot_2018, Abbot_2019} from Binary Neutron Star Merger (BNSM) has helped in constraining the EoS using the tidal deformability measurements \cite{Raithel_2018, De}. However, an abundance of mysteries is expected to unfold once we detect the emissions from the post-merger phase \cite{Sarin_2021}. Numerical Relativity (NR) simulations have helped us gain insights into the possible merger and post-merger dynamics \cite{Baiotti_2017, Baiotti_2019, Radice_2020, Dietrich_2021,Bauswein_2010_2,Bauswein_2012_2,Ecker_2025,oechslin_2003,oechslin_2004,Oechslin_2002}. A substantial amount of work has studied the behaviour of BNSM systems depending on the EoSs~\cite{Raithel_2022_2,Takami_2017, gw_ana,gth,Takami_2014, Jocelyn_2013,Tootle_2021,Sekiguchi_2011_2,Vijayan_2023,Bauswein_2012,Bauswein_2014,Bauswein_2015}, and microphysics~\cite{Raithel_2022, Most_2022, Most_2020_2, Most_2023, Hanauske_2019,Bruno_2011,Rezzolla_2011,Kenta_2014,Sekiguchi_2011,Kenta_2018,Foucart_2018,Foucart_2020,Blacker_2023,Blacker_2023_2,Just_2023,Henrique_2019,Kiuchi_2024,Carlos_2015,Anderson_2008}. BNSM events are rich systems that can help us probe the PT process. A significant line of investigations has accounted for hadron-quark PT in such systems~\cite{Takami_2022,Hanauske_2019_2, Most_2019,Most_2020_3,Weih_2020,Tootle_2022,Kedia_2022,Aviral_2021,Fujimoto_2023,Haque_2023, Ujevic_2023,Fujimoto_2025,Bauswein_2020,Bauswein_2020_2,Bauswein_2019,Blacker_2020}.

PT at intermediate densities can give rise to a new class of compact objects called the hybrid stars (HSs) having pure quarks or hadron-quark mixed phase in the core~\cite{Alford_2005, Nandi:2017rhy, Nandi:2018ami, Nandi_2021, Mallick:2021jgb, Anil, Li_2023}. The EoSs having PT in NSs have been popularly categorized between two extremes--- $(i)$ Maxwell construction , having a density discontinuity at a constant pressure, $(ii)$ Gibbs construction, where the pressure varies in the mixed phase where hadrons and quarks coexist \cite{Contrera_2022, Qin_2023,Sen_2024, Brandes_2024,Han_2019,Rather_iop,Glendening}. However, the characteristics of PT are not known a priori. It depends on the surface tension between the two adjacent phases of fluid--- hadron and quark \cite{wu_finite-size_2017,Constantinou_2023}. Such features 
have been studied using the pasta phase \cite{Maslov_2019,Ju_2021,Thi,Voskresensky_2002,TATSUMI_2003,Tomoki_2006,Noda_2013,Yasutake_2014}, taking into account the surface tension and Coulomb effects arising due to different shapes and structures of elements in mixed phases. Several studies have extended the conventional Maxwell construction by incorporating modifications that emulate the outcomes of pasta matter to understand its imprints in the characteristics of HSs \cite{Ayriyan2017,Abgaryan:2018,Blaschke_2020,Alvarez-Castillo-Blaschke,Shahrbaf, Ayriyan2017,Pereira_2022,Pradhan_2024}. These studies have not yet considered the additional complexities that arise in BNSM, especially the post-merger dynamics. The presence of PT can affect the GW signal and the post-merger peak in the power spectral density (PSD) \cite{Most_2019, Blacker_2023_2, Weih_2020, Bauswein_2019}.  However, all these studies either assume a Maxwell or a Gibbs type of PT. This article presents the first attempt to explore the mixed phases that lie between these two extremes. Our analysis shows how the post-merger signal can help us identify the nature and type of PT.
\begin{figure*}[ht!]
    \centering
    \includegraphics[scale=1]{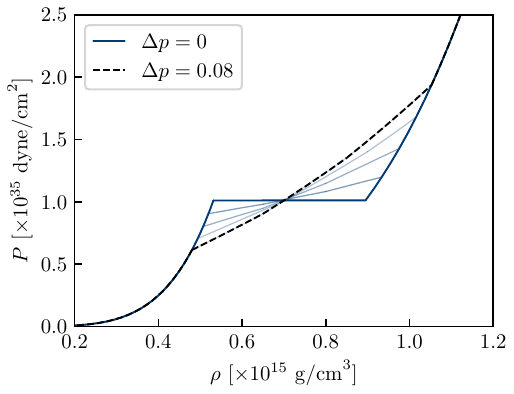}
    \includegraphics[scale=1]{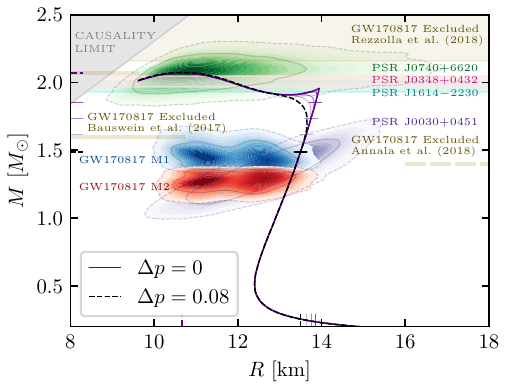}

    \caption{ [Left] The pressure vs rest-mass density plot of the EoS set. [Right] The mass--radius ($M$--$R$) sequences corresponding to the EoS set with varying $\Delta p$. The ticks on the y-axis (x-axis) denote the masses (radii) where central rest-mass density enters the mixed phase regime of that EoS. The purple dashed tick on the y-axis (x-axis) marks the maximum mass (and its radius). Current astrophysical constraints are satisfied~\cite{ozel_2010,Antoniodis_2013,GW170817,Abbot_2018,Abbot_2019,Miller_2019,Riley_2019,Miller_2021,Riley_2021,Annala_2018,Bauswein_2017_2,Rezzolla_2018_2,Lattimer_1990,Lattimer_2001}.}

\label{fig:ppeos_plot}
\end{figure*}
We aim to understand PT with a control parameter $\Delta p$ that can be varied given a pair of fixed hadronic and quark EoS. It allows us to survey a range of EoS starting with a first-order phase transition (FOPT) corresponding to a Maxwell construction ($\Delta p=0$) between hadronic and quark EoS and increase the $\Delta p$ to the point that approaches Gibbs construction ($\Delta p\approx0.08$)~\cite{Maslov_2019}. Implementing these EoSs in merger simulations is of utmost interest to reveal the direct impact of different phases on crucial observational emissions like GW and ejecta mass. From our simulations, we report specific signatures that exclusively depend on the type of PT. If such signatures are found in future multi-messenger observations, we can remarkably constrain the EoS of matter at this regime. We report the first correlation of the threshold mass for prompt collapse with varying $\Delta p$ in light of GW170817.


\section{Formalism}\label{sec:form}
We construct the hybrid EoS using piecewise polytropes similar to Refs \cite{Alvarez-Castillo-Blaschke, Abgaryan:2018, Blaschke_2020,Pereira_2022, Shahrbaf, Ayriyan2017}. Our construction of EoS begins above the nuclear saturation density ($n_{0} = 0.16~\mathrm{fm}^{-3}$), and the pressure is defined as, $P(n) = \kappa_{i} (n/n_{0})^{\Gamma_{i}}, \:  n_{i}<n<n_{i+1}, \: i=1...4$. The polytropic index is defined as $\Gamma_{i}$, where $i$ denotes the segment, $n$ denotes the number density, and $\kappa_i$ is the polytropic factor. Table 1 of Ref.~\cite{Abgaryan:2018} gives the parameters for the four polytropes. The second segment corresponding to $\Gamma_{2} = 0$ denotes a density jump corresponding to Maxwell's construction. This EoS uses a replacement interpolation method (RIM) to obtain the EoSs shown in Fig.~\ref{fig:ppeos_plot} (left). We have constructed a set of 5 EoSs with varying $0 \leq \Delta p \leq 0.08$, keeping the hadronic and quark EoS fixed. This EoS set is compatible with the current observational constraints extracted from NICER and LIGO-Virgo observations~\cite{ozel_2010,Antoniodis_2013,GW170817,Abbot_2018,Abbot_2019,Miller_2019,Riley_2019,Miller_2021,Riley_2021,Annala_2018,Bauswein_2017_2,Rezzolla_2018_2,Lattimer_1990,Lattimer_2001}. 

In the merger setup, the EoSs are re-established in the formalism described in Refs~\cite{adm13,ppeos}. We note that the pressure in the PT regime is supposed to be constant with the change in density for EoS-$\Delta p_{0.0}$, which can be established by setting $\Gamma=0$ in piecewise-polytropic EoS implementation. However, the speed of sound $c_\mathrm{s}^2=0$ in this density regime causes issues in numerical hydrodynamic evolutions~\cite{Espino_2022,Font_1994,adm13}. We avoid this issue by setting $\Gamma=0.01$. This value, being non-zero, is enough to introduce a thin numerical mixed phase. The thermal part is approximately supplemented using an assumption of ideal-gas behaviour with thermal index $\Gamma_\mathrm{th}=1.75$~\cite{gth,gth2}. The temperature effects due to a change in the degrees of freedom around the PT (and reverse PT) are not taken into account. Therefore, the evolving system may not keep the PT process adiabatic. We used the \textsc{WhiskyTHC} code~\cite{thc1,thc2,thc3} for solving the General-Relativistic Hydrodynamic (GRHD) equations \cite{grmhd}. More information about the formalism is described in the Appendix.

\section{Results} \label{sec:Results}

\begin{figure}
    \centering
    \includegraphics[scale=0.95]{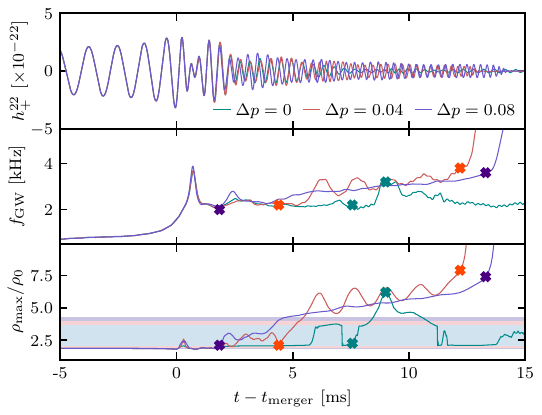}
    \caption{Evolution of the intermediate mass merger. [Top] Extraction of $h_+^{22}$ strain at 100 Mpc distance from the merger. [Middle] Evolution of $f_\mathrm{GW}$ marked with the timestamps in alignment with time stamps marked in $\rho_\mathrm{max}$ evolution. [Bottom] Evolution of $\rho_\mathrm{max}$ (in terms of $\rho_0$ = $2.51 \times 10^{14}~\mathrm{g/cm^3}$~\cite{glen}) marked with the start and end of the timespan of our interest. The shaded region indicates the region between the end of pure HM and the beginning of pure QM in our EoSs.}
    \label{fig:1.36_evolution}
\end{figure}

In this work, we investigated the delayed PT and PT-triggered collapse scenarios~\cite{Weih_2020}. The NSs involved in these mergers have central rest-mass density below the onset of PT; hence, PT is experienced by the matter only during (or after) the merger. Figure~\ref{fig:1.36_evolution} monitors the evolution of the post-merger remnant resulting from intermediate equal mass mergers ($2.72~M_\odot$). A detailed investigation of low-mass and heavy-mass mergers is discussed in Appendix D and E, respectively. The maximum rest-mass density ($\rho_\mathrm{max}$) in Fig.~\ref{fig:1.36_evolution} contains a shaded region in the background, indicating the density range corresponding to the mixed phase. Since the lower boundary of this shaded region marks the end of the pure hadronic phase, and the upper boundary marks the beginning of the pure quark phase, the shaded region helps in tracking whether the post-merger remnant has pure HM or hybrid matter at a particular instant. The markers indicate the part of the evolution which are of interest to us. These markers are used again in subsequent plots to focus on the exact time stamp marked in $\rho_\mathrm{max}$ evolution. In the mass range considered in work, the post-merger remnant with pure HM ($\rho_\mathrm{max}$ below the onset density of PT) is identified as hypermassive NS (HMNS) \cite{Sarin_2021,Metzger_2019}. The post-merger remnant that has mixed phase or pure QM at core ($\rho_\mathrm{max}$ above the onset density of PT) is identified as hypermassive HS (HMHS) \cite{Hanauske_2021}. 

\begin{figure*}
    \centering
    \begin{minipage}{0.95\textwidth}
    \centering\hspace*{-1.7cm}
        \includegraphics[scale=0.9]{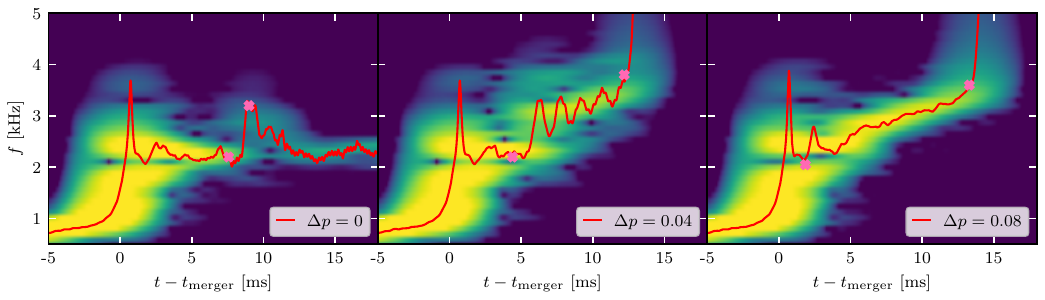}
    \end{minipage}
    \begin{minipage}{0.01\textwidth}
    \centering\vspace*{-0.5cm}\hspace*{-1.5cm}
        \includegraphics[scale=0.9]{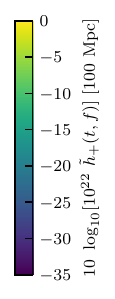}
    \end{minipage}
    
    \caption{Spectrogram of the GW signal from the intermediate mass merger. The red curve shows $f_\mathrm{GW}$ for that particular EoS, marked with time stamps from Fig.~\ref{fig:1.36_evolution}.}
    \label{fig:1.36_spectogram}
\end{figure*}
\begin{figure}
    \centering
    \includegraphics[scale=1]{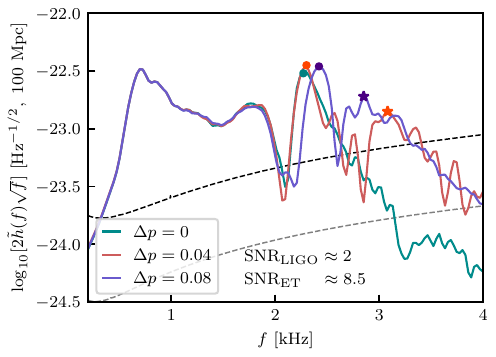}
    \caption{PSD of GW signals from the intermediate mass merger at 100 Mpc. The $f_2^h$ and $f_2^q$ peaks are indicated with circles and star markers, respectively. The black and grey dashed lines are sensitivity curves of LIGO A+~\cite{Capote_2025} and ET-D~\cite{Hild_2011}, for which the post-merger signal-to-noise ratio (SNR) for each $\Delta p$ is given in Table~\ref{tab:snr} in the Appendix.}
    \label{fig:1.36_psd}
\end{figure}

For the EoS-$\Delta p_{0.08}$, we observe a gradual increase in $\rho_\mathrm{max}$, which leads to the introduction of {pure QM}\footnote{Note that this applies only to the particular region, primarily around the point where the density approaches $\rho_\mathrm{max}$ in the remnant. Most of the outer part of the remnant still remains in the hadronic phase.} at $\sim 5~\mathrm{ms}$. The instantaneous frequency $f_{\mathrm{GW}}$ (see Appendix C) also increases gradually. It confirms that the steady increase in frequency results from HMNS converting to HMHS. After that, the $\rho_\mathrm{max}$ enters exponential growth, indicating a collapse into BH. This signature is also reflected in the spectrogram plotted in Fig.~\ref{fig:1.36_spectogram} (right panel).

For the case of $\Delta p_{0.04}$, the transition of the hypermassive remnant from HMNS to HMHS took a shorter time, leading to a steeper growth in $\rho_\mathrm{max}$. However, the HMHS core experiences numerous bounces in the form of density compressions and expansions, after which the remnant enters an exponential growth in $\rho_\mathrm{max}$ due to a collapse into BH. The bouncing characteristic of $f_{\mathrm{GW}}$ leaves a broader frequency spread in the spectrogram between the two markers.

For EoS-$\Delta p_{0.0}$, the $\rho_\mathrm{max}$ evolution shows that the post-merger remnant contained pure HM for a longer period than other cases. The remnant converts from HMNS to HMHS at $\sim 9~\mathrm{ms}$. However, the HMHS state is short-lived and immediately experiences a reverse PT~\cite{Ujevic_2023} at $\sim 11~\mathrm{ms}$, converting back to the HMNS. In this case, the transition between hadronic EoS and quark EoS is the softest. Hence, the transition from HMNS to HMHS is triggered by a mini-collapse, which appears as an exponential-like growth in the $\rho_\mathrm{max}$ evolution that stalls momentarily in the pure quark regime. When the incoming matter hits the sufficiently stiff core, the recoil is so high that the rest-mass density reduces rapidly inside the core, leading to a reverse PT. This phenomenon takes place in $\sim 1~\mathrm{ms}$ timescale, resulting in an immediate spike in $f_\mathrm{GW}$. This spike is also reflected in Fig.~\ref{fig:1.36_spectogram} (left) above the saturated value (associated with HMNS state) of post-merger frequency, precisely at the time stamp where the post-merger remnant entered the HMHS state. Such signatures in future detection can be a direct signature of a sharp FOPT. The appearance of this signature has been consistently observed in low-mass and heavy-mass mergers (see Figs.~8 and 11 in the Appendix). After returning to the HMNS state, we observe the $\rho_\mathrm{max}$ re-entering the shaded region at $\sim 14~\mathrm{ms}$.

In Fig.~\ref{fig:1.36_psd}, we observe the PSD of the GW emissions from these systems at 100 Mpc. We identified distinct $f_2$ peaks appearing from two phases of the post-merger remnant. We mark the frequency peak as $f_2^h$, which is caused by the remnant being in the HMNS state. We found that $f_2^h$ frequencies increase with an increase in $\Delta p$. For the case of EoSs---$\Delta p_{0.08}$ and $\Delta p_{0.04}$, we observe an extra peak denoted as $f_2^{q}$. These peaks are expressed due to the conversion of the post-merger remnant from the HMNS state to the HMHS state. Since the remnant formed by EoS-$\Delta p_{0.0}$ remains in the HMNS state, such peaks did not appear in the PSD. The observation of such peaks in future detectors will confirm the evidence of hadron-quark PT. The $f_2^{q}$ has been previously shown to be associated with delayed PT~\cite{Weih_2020}. We found evidence of signature being triggered by PT and showed concrete dependence of the peak frequency on the $\Delta p$, that is, the $f_2^{q}$ frequency decreases with an increase in $\Delta p$. This signature has been consistently observed in low-mass and heavy-mass mergers (see Figs.~7 and 10 in the Appendix).

The threshold mass ($M_{\mathrm{thres}}$) for prompt collapse is important to quantify in the context of post-merger dynamics and has been shown to be directly dependent on the characteristics of EoS~\cite{Bauswein_2021,Bauswein_2020_3,Bauswein_2020_2,Hotokezaka_2011,Koppel_2019,Bauswein_2013,Ecker_2025_2}. We establish the dependence of $M_{\mathrm{thres}}$ on the parameter $\Delta p$. We simulated varying equal mass mergers and extracted the final state of the remnant at $\sim 50~\mathrm{ms}$ of evolution. There exists a minimum $\Delta p$ in the EoS for a fixed total mass, that will trigger the BH collapse. The change in state (from HMNS/HMHS to BH) of the final remnant has been presented in Table~\ref{tab:threshold_mass}. We find that the relation between $M_{\mathrm{thres}}$ and $\Delta p$ can be fit using a hyperbolic tan function, 
\begin{equation}\label{eq:m_thres}
    M_\mathrm{thres}(\Delta p) = a \tanh{\{b(\Delta p+c)\}} + d,
\end{equation} where the fitting parameters are: $a=-0.03~M_\odot$, $b=10.00$, $c=-4.00$, $d=2.72~M_\odot$. For such a fitting we find the maximum fractional percentage error \cite{chatterjee_insights_2025} to be $\lesssim 2.23\%$. The fitting function used in our study may not universal and only an attempt to show a correlation between $\Delta p$ and $M_\mathrm{thres}$. To understand this correlation with further clarity, a larger campaign of simulations with a larger ensemble of EoSs is required. $M_{\mathrm{thres}}$ also depends on many other factors, these fitting coefficients will change if a different combination of hadronic and quark EoS is used to make a new set of EoSs. Mainly, a stiffer EoS set will support heavier NSs, leading to higher $M_\mathrm{thres}$. This will directly increase the value of parameter $d$. This result can also be affected by changes in mass ratios in unequal mass mergers \cite{Tootle_2022,Fujimoto_2025,Bauswein_2020_3}.

From Table~\ref{tab:threshold_mass} we observe that, for the mass range $1.35$--$1.37~M_\odot$, the merger remnant will be long-lived or experience delayed collapse if $\Delta p \lesssim 0.04$. GW170817 has been predicted to have formed a long-lived remnant or have experienced delayed collapse into BH \cite{Piro_2018,Gill_2019,Sarin_2021,Margalit_2017,Yu_2018,Ai_2018,Berthier_2017,Agathos_2020,Rezzolla_2018}. In the regime of our EoS, we can conclude that such a scenario would require the $\Delta p \lesssim 0.04$. However, an EoS with $\Delta p \gtrsim 0.04$ can still be ruled in, if GW170817 is believed to have a prompt collapse into BH or if an EoS set with a higher $M_\mathrm{TOV}$ is considered.

\begin{table*}
	\caption{The table indicates the change in remnant fate for the equal mass merger.}
	\centering
 \def\arraystretch{1.2}
	\begin{tabular}{@{\hspace{0.5cm}}c@{\hspace{0.5cm}}c@{\hspace{0.5cm}}c@{\hspace{0.5cm}}c@{\hspace{0.5cm}}c@{\hspace{0.5cm}}c@{\hspace{0.5cm}}c@{\hspace{0.5cm}}c@{\hspace{0.5cm}}c@{\hspace{0.5cm}}}
		\hline
		  $\Delta p$ & 2.68~$M_\odot$ & 2.69~$M_\odot$ & 2.70~$M_\odot$ & 2.71~$M_\odot$ & 2.72~$M_\odot$ & 2.73~$M_\odot$ & 2.74~$M_\odot$ & 2.75~$M_\odot$ \\
		\hline
            0 & - & - & - & - & - & - & - & BH \\
            0.02 & - & - & HMHS & HMHS & HMHS & HMHS & HMHS & - \\ 
            0.04 & - & - & BH & BH & BH & BH & BH & - \\ 
            0.06 & - & HMHS & - & - & - & - & - & - \\
            0.08 & HMHS & BH & - & - & - & - & - & - \\ 
		\hline
	\end{tabular}
	\label{tab:threshold_mass}
\end{table*}

\section{Summary and Conclusion} \label{sec:Summ}

Using numerical relativity simulations, we present the possible signatures that can pinpoint the nature of the hadron-quark phase transition in binary neutron star mergers. We have identified distinct peak frequencies $f^q_2$ in the PSD that are entirely attributed to the PT of the post-merger remnant. Considering the signal-to-noise ratio (SNR) detectability threshold for BNS being $\sim8$, the signal has promising detectability in the third generation detectors \cite{Mould_2024,Petrov_2022,Iacovelli_2022,Ronchini_2022}. The $f^q_2$ frequencies are highly dependent on the $\Delta p$ of EoS; that is, $f^q_2$ decreases with the increase in $\Delta p$. Complementing this signature, the nature of PT leaves specific imprints on the spectrogram. The Maxwell-type PT leaves a spike in $f_\mathrm{GW}$ during the PT, which is also reflected in the spectrogram above the $f^{h}_2$ frequency associated with the HMNS state. A Gibbs-like PT shows a gradual growth in frequency. This forms a robust two-folded signature from gravitational wave analysis. These signatures have been consistently observed across varying total mass (see the analysis of low-mass and high-mass mergers in Appendix D and E, respectively), different mass ratios (see Appendix H), and different choices of hadronic and quark EoSs (see Appendix G).

We establish a correlation between the threshold mass for prompt collapse and the parameter $\Delta p$, which is fit using a hyperbolic tangent function with suitable parameters. Our analysis shows that $\Delta p \lesssim 0.04$ is required if GW170817 formed a long-lived remnant or has experienced a delayed collapse into a black hole.

We observed the effect of $\Delta p$ on the ejecta mass from the event (see Appendix F). However, the effects are dissimilar for low-mass and heavy-mass mergers, indicating that the end fate of the remnant brings additional complexity to the study~\cite{Bauswein_2013_2,Lehner_2016,Radice_2018_2,Ciolfi_2017}. Further investigation on the direct impact on kilonova afterglow and nucleosynthesis process is needed.

Our EoS model uses piecewise polytropes, which is useful to draw the proof-of-concept scenario. However, future studies employing realistic microphysical temperature-dependent EoS are expected to shed more comprehensive insights \cite{Blacker_2023_2}. Since these distinct signatures from the post-merger GW signal appear within a few milliseconds of merger, we expect them to be consistently expressed but may change quantitatively.

\section*{Acknowledgement}

The authors are grateful to Fernando Navarra, David Radice, Spandan Sarma, J\'er\^ome Novak, Asim K.~Saha, Atul Kedia, and Skund Tewari for the helpful discussions and simulation setup. SH and RM acknowledge the SERB grant: CRG/2022/000663. SC acknowledges the PMRF scheme by the MoE, Govt.~of India and support from the Ministry of Education and Research, Romania, CNCS/CCCDI-UEFISCDI, Project No.~PN-IV-P1-PCE-2023-0324. KKN acknowledges the DAE project: RIN4001-SPS. Simulations were performed at the clusters--- KALINGA (NISER) and PARAM Ganga (IIT Roorkee, NSM resource by C-DAC and supported by MeitY and DST). The data-handling and post-processing is done using the following packages--- {\sc NumPy}~\cite{numpy}, {\sc SciPy}~\cite{scipy}, {\sc Matplotlib}~\cite{matplotlib}, {\sc Seaborn}~\cite{Waskom_2021}, {\sc Pandas}~\cite{mckinney_2010}, {\sc Kuibit}~\cite{kuibit1,kuibit2}, {\sc Jupyter}~\cite{jupyter}.\\

\noindent SC and SH contributed equally to this work.

\appendix
\section{Equation of State}

\begin{table*}
    \caption{ Table of piecewise polytropic parameters that were used in simulations. Four pieces are used for crust EoS \cite{ppeos} in addition to the number of pieces used for respective EoSs reported below.}
	\centering
	\begin{tabular}{c@{\hspace{0.5cm}}c@{\hspace{0.5cm}}c@{\hspace{0.5cm}}c@{\hspace{0.5cm}}c@{\hspace{0.5cm}}c@{\hspace{0.5cm}}c@{\hspace{0.5cm}}c@{\hspace{0.5cm}}c@{\hspace{0.5cm}}c@{\hspace{0.5cm}}c@{\hspace{0.5cm}}c@{\hspace{0.5cm}}c}
		\hline
		  Polytrope Pieces & \multicolumn{2}{c}{EoS-$\Delta p_{0}$} & \multicolumn{2}{c}{EoS-$\Delta p_{0.02}$} & \multicolumn{2}{c}{EoS-$\Delta p_{0.04}$} & \multicolumn{2}{c}{EoS-$\Delta p_{0.06}$} & \multicolumn{2}{c}{EoS-$\Delta p_{0.08}$} \\
            $i$ & $\Gamma_i$ & $\log\rho_i$ & $\Gamma_i$ & $\log\rho_i$ & $\Gamma_i$ & $\log\rho_i$ & $\Gamma_i$ & $\log\rho_i$ & $\Gamma_i$ & $\log\rho_i$ \\
		\hline
            1 & 4.921 & 14.3135 & 4.921  & 14.3135 & 4.921 & 14.3135 & 4.291 & 14.3135 & 4.921 & 14.3135 \\
            2 & 0.01  & 14.7256 & 0.36   & 14.7157 & 0.675 & 14.7051 & 1.04  & 14.6938 & 1.28 & 14.6815 \\
            3 & 4.0   & 14.9518 & 0.478  & 14.8129 & 0.92  & 14.8129 & 1.29  & 14.8451 & 1.5   & 14.8129  \\
            4 & 2.8   & 15.099  & 0.6546 & 14.903  & 1.1   & 14.903  & 1.49  & 14.9542 & 1.677 & 14.9294 \\
            5 &       &         & 4.0    & 14.97   & 4.0   & 14.9892 & 4.0   & 15.0067 & 4.0   & 15.0222 \\
            6 &       &         & 2.8    & 15.099  & 2.8   & 15.099  & 2.8   & 15.099  & 2.8   & 15.099  \\ 
		\hline
	\end{tabular}
	\label{ppeos_table}
\end{table*}

We construct the hybrid EoS using piecewise polytropes. This work uses four segments of polytropes similar to Refs \cite{Alvarez-Castillo-Blaschke, Abgaryan:2018}, called ``ACB4". Our construction of EoS begins above the nuclear saturation density ($n_{0} = 0.16~\mathrm{fm}^{-3}$). The polytropic index is defined as $\Gamma_{i}$, where $i$ denotes the segment, $n$ denotes the number density, and $\kappa_i$ is the polytropic factor. The parameters for the four polytropes are given in Table 1 of Ref.~\cite{Abgaryan:2018}. The second segment corresponding to $\Gamma_{2} = 0$ denotes a density jump corresponding to Maxwell's construction. The pressure can be reformulated in terms of baryon chemical potential ($\mu$) and can be written as, 
\begin{equation}\label{eq2}
    P(\mu) = \dfrac{\kappa_{i}}{n_{0}^{\Gamma_{i}}} \Big[\left(\mu - m_{0,i} \right) \dfrac{\Gamma_{i}-1}{\kappa_{i}\Gamma_{i}}n_{0}^{\Gamma_{i}} \Big]^{\Gamma_{i}/(\Gamma_{i}-1)},
\end{equation}
where the effective masses of the constituent degrees of freedom in each phase are denoted by $m_{0,i}$. The EoS constructed in this method also fulfils the minimum stability criteria of PT called the Seidov limit \cite{Seidov_1971}.  We obtain the relation between chemical potential and number density,
\begin{equation}\label{eq3}
    \mu = \left(\dfrac{\kappa_i \Gamma_i}{ \Gamma_{i}-1}\right) \left(\dfrac{n^{\Gamma_{i}-1}}{n_{0}}\right) + m_{0,i}.
\end{equation} 
The energy density at every point is given by $\epsilon = -p + \mu n$. The RIM interpolation scheme uses a polynomial function for generating the EoSs \cite{Abgaryan:2018, Pereira_2022}. The pressures at the hadronic and quark phases are denoted by $P_H (\mu)$ and $P_Q (\mu)$, respectively. The phase in between them is called the mixed phase (except in a scenario where we have Maxwell construction), and can be given by the polynomial function,
\begin{equation} \label{eq5}
    P_M (\mu) = \alpha (\mu - \mu_c)^p + \beta (\mu-\mu_c)^q + (1+\Delta p) P_c.
\end{equation}
where $\mu_c$ is called the critical chemical potential at which the first-order phase transition occurs, and we have the pressures of both the phases equal ($P_Q(\mu_c) = P_H(\mu_c) = P_c$). In this work we take only the quadratic form $(p=2,~q=1)$ of \cref{eq5}, which reduces it to,
\begin{equation} \label{eq6}
    P_M (\mu) = \alpha (\mu - \mu_c)^2 + \beta (\mu-\mu_c) + (1+\Delta p) P_c.
\end{equation}
The boundary conditions corresponding to these sets of equations demand a smooth transition from the hadronic to the quark phase. Hence, the pressures at each of these phases should be equal to the pressure at the mixed phase, and the number density should also match the number densities at the mixed phase. From the boundary conditions, we can solve for the values of $\alpha$ and $\beta$. For detailed description see Refs.~\cite{Alvarez-Castillo-Blaschke,Abgaryan:2018, Pereira_2022,Shahrbaf, Ayriyan2017}.

\begin{table}
	\caption{Table of SNR for post-merger GW signal for various $\Delta p$ in Fig.~\ref{fig:1.36_psd}:}
	\centering

	\begin{tabular}{c@{\hspace{0.5cm}}c@{\hspace{0.5cm}}c}
		\hline
		  $\Delta p$ & LIGO A+~\cite{Capote_2025} & ET-D~\cite{Hild_2011} \\
		\hline
            0 & 1.91 & 8.13 \\
            0.04 & 2.01 & 8.55 \\ 
            0.08 & 2.06 & 8.72 \\ 
		\hline
	\end{tabular}

	\label{tab:snr}
\end{table}

We have constructed a set of 5 EoSs with varying $0 \leq \Delta p \leq 0.08$, keeping the hadronic and quark EoS fixed. 
We mimic the EoSs using piece-wise polytrope formalism~\cite{adm13,ppeos}, which requires breaking a tabulated EoS into $N$ pieces of density ranges. For each piece $i$ (range $\rho_{i}\leq\rho<\rho_{i+1}$), we fit the curve in function $P=K_i\rho^{\Gamma_i}$, where ($K_i,\Gamma_i$) are the $i^\mathrm{th}$ polytropic constant and polytropic exponent respectively. After fitting a reasonable $\Gamma_i$, the next polytropic constant $K_{i+1}$ is determined using the boundary condition to ensure the smoothness of the mimicked EoS. We perform the piece-wise polytropic fit to all the EoSs. The final EoS set used in the simulations is given in \cref{ppeos_table}.

During the merger evolution, the EoSs are supplemented by an ideal-fluid thermal component~\cite{gth2}, which accounts for shock heating in the system that dissipates the kinetic energy into the internal energy. The thermal component $\Gamma_\mathrm{th}$ is set to 1.75~\cite{gth}. This comes with a caveat that the thermal part is approximate and does not accurately reproduce the effects of temperature at a high-density regime given by the finite-temperature EoSs \cite{Bauswein_2010,Raithel_2021, Figura_2020}, and the behaviour of PT could also be affected by the rise in local temperatures.

\section{Simulation Setup}\label{subsec:simulation_setup}

We used the \textsc{WhiskyTHC} code~\cite{thc1,thc2,thc3} for solving the General-Relativistic Hydrodynamic (GRHD) equations \cite{grmhd}. 
These equations are defined in a conservative form using finite-difference method with high-resolution shock capturing (HRSC) scheme~\cite{hrsc1,hrsc2}. A fifth-order accurate monotonicity-preserving (MP) MP5~\cite{mp5_1,mp5_2} scheme is used as the flux reconstruction method along with standard Harten-Lax-van Leer-Einfeldt (HLLE) approximate Riemann solver~\cite{hlle1,hlle2}. The spacetime is evolved using the Z4c formalism~\cite{z4c1,z4c2}, which is implemented through {\sc CTGamma} code~\cite{ctg1,ctg2} built in the {\sc Einstein Toolkit}~\cite{ET2,ET3}. It is based on the \textsc{Cactus Computational Toolkit}~\cite{cactus,cactus2}, a software framework designed with \textsc{Carpet} adaptive mesh refinement (AMR)~\cite{carpet1,carpet2,amr} driver for high-performance computing. {\sc CTGamma} uses fourth-order finite-differencing with a fifth-order Kreiss-Oliger~\cite{Kreiss_1973} artificial dissipation to ensure non-linear stability of the evolution. Method-of-lines handles the spacetime and hydrodynamics coupled evolution using a stability-preserving third-order Runge-Kutta scheme~\cite{Gottlieb_2009} for time integration.

The simulations carried out for this work used the coarsest grid size of $\Delta x = 10 $~$M_\odot$ (with 7 refinement levels), having the finest resolution resolving the star to be $\Delta x = 0.156 $~$M_\odot$ ($\sim 230~\mathrm{m}$). An extra refinement level is added to handle the post-merger remnant at $\sim 115~\mathrm{m}$ resolution, located at the centre. The total extent of the domain is $1000~M_\odot$ ($\sim 1470$~km), and a plane symmetry is applied with respect to the $z = 0$ plane. The resolution setup used in this work has been found to be acceptable in numerous studies in NR simulation of BNSM systems~\cite{parma1,Takami_2014,Bernuzzi_2016,Weih_2020,Hanauske_2021,Kedia_2022,Takami_2022,Ujevic_2023}.

The initial configuration data for our simulations are generated using the \textsc{Lorene} library~\cite{lorene1,lorene2}. It uses multi-domain spectral methods to solve the partial differential equations. These data are obtained using the assumptions of quasi-circular equilibrium in the coalescence state and conformally flat metric to solve the conformal thin-sandwich equations~\cite{lorene3}. In this work, we have set the initial physical separation between the stars to be $45~\mathrm{km}$ with irrotationality of the fluid flow, which is defined as employing vanishing vorticity. It allows $\sim3$ orbits before merging for the equal mass system of $2.72~M_\odot$. Using the estimation technique described in \cite{Kyutoru_2014}, the residual eccentricity is found to be $\sim0.025$.

\section{GW extraction}

Here, we describe the method of extracting gravitational waveforms from our simulations in detail. First, we extract the Weyl scalar (particularly $\Psi_4$) from the simulations using the Newman-Penrose formalism~\cite{np}. The complex variable $\Psi_4$ provides a measure of outgoing radiation and can be related to the complex GW strain $h$ by its second-order time differentiation~\cite{hdotdot},  
\begin{equation}
    \ddot{{h}}=\ddot{h}_{+}-i\ddot{h}_\times=\Psi_4,
\end{equation}

where $h_+$ and $h_\times$ are the polarisation modes, which can be decomposed using spin-weighted spherical harmonics~\cite{spin} of spin weight $-2$,
\begin{align}
	\Psi_4(t,r,\theta,\phi)=\sum_{l=2}^{\infty}\sum_{m=-l}^{l}\Psi^{lm}_4(t,r)_{-2}Y^{lm}(\theta,\phi).
\end{align} 
We have analysed the dominant mode $l=m=2$ to find the ${h}^{22}$ strain (at 100 Mpc) by double integrating $\Psi^{22}_4$ over time numerically. The expected event rate at $\mathrm{100~Mpc}$ is 0.04--7.12 $\mathrm{yr}^{-1}$ approximately~\cite{Abbott_2023}. Now, $h^{22}$ being a complex variable,
\begin{equation}
    h^{22}=h^{22}_+ -ih^{22}_\times = |h^{22}|e^{i\phi},
\end{equation}
where $|h^{22}|$ is the GW amplitude, and $\phi$ is the phase of the complex waveform. We have set the \textit{merger time} as the point where  $|h^{22}|$ is maximum. In our results, we have measured $t-t_{\textrm{merger}}$ in our time axes and aligned all the GW waveforms at $t-t_{\textrm{merger}}=0$. The \textit{instantaneous frequency} is calculated by,
\begin{equation}
    f_\mathrm{GW}=\frac{1}{2\pi}\frac{d\phi}{dt}.
\end{equation}

We have calculated the PSD of the GW amplitude, which is defined as,
\begin{equation}
    \Tilde{h}(f) =  \sqrt{\frac{|\Tilde{h}_+(f)|^2+|\Tilde{h}_\times(f)|^2}{2}},
\end{equation}
where $\Tilde{h}_{+,\times}(f)$ are the Fourier transforms of ${h_{+,\times}(t)}$ given as,
\begin{equation}
   \tilde{h}_{+, \times}(f) \equiv \begin{cases}\int h_{+, \times}(t) e^{-i 2 \pi f t} d t, & f \geq 0 \\ 0, & f<0.\end{cases}
\end{equation}

The SNR is computed as,

\begin{equation}
    \mathrm{SNR} \equiv \left[ \int^\infty_0 \frac{|2\tilde{h}(f)\sqrt{f}|^2}{S_h(f)} \frac{df}{f} \right]
\end{equation}
where $S_h(f)$ is the noise PSD of the detector. For an approximate calculation of SNR for the $f^{h}_2$ and $f^{q}_2$ peaks in LIGO A+~\cite{Capote_2025} and ET-D~\cite{Hild_2011}, we used the GW post-merger signal ($ t - t_{\mathrm{merger}} > 1.5~\mathrm{ms}$).

\begin{figure}
    \centering
    \includegraphics[scale=0.95]{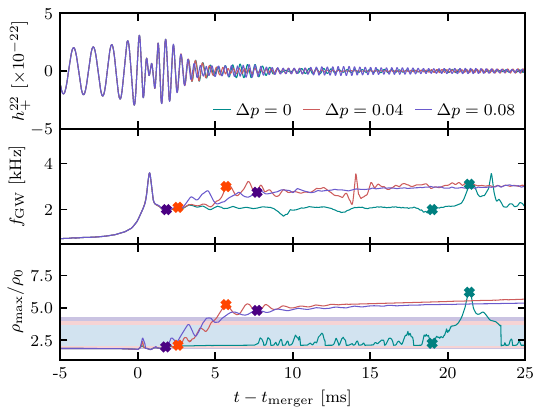}
    \caption{Evolution of the low mass merger. [Top] Extraction of $h_+^{22}$ strain at 100 Mpc distance from the merger. [Middle] Evolution of $f_\mathrm{GW}$ marked with the timestamps in alignment with time stamps marked in $\rho_\mathrm{max}$ evolution. [Bottom] Evolution of $\rho_\mathrm{max}$ marked with the start and end of the timespan of our interest. The shaded region hints at the mixed-phase region for the particular EoS.}
    \label{fig:1.34_evolution}
\end{figure}
\begin{figure}
    \centering
    \includegraphics[scale=1]{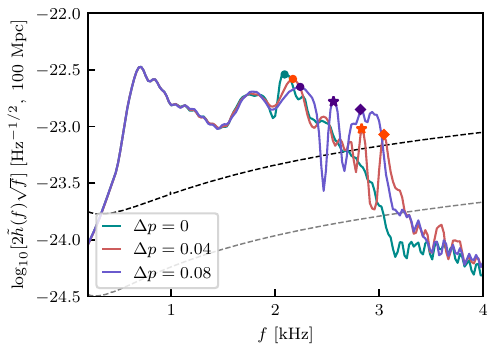}
    \caption{PSD of GW signals from low-mass merger at 100 Mpc. The $f_2^h$, $f_2^{q1}$ and  $f_2^{q2}$ peaks are indicated with circle, star and diamond markers, respectively. The black and grey dashed lines are sensitivity curves of LIGO A+~\cite{Capote_2025} and ET-D~\cite{Hild_2011}.} 
    \label{fig:1.34_psd}
\end{figure}

\begin{figure*}
    \centering
    \begin{minipage}{0.95\textwidth}
    \centering\hspace*{-1.7cm}
        \includegraphics[scale=0.9]{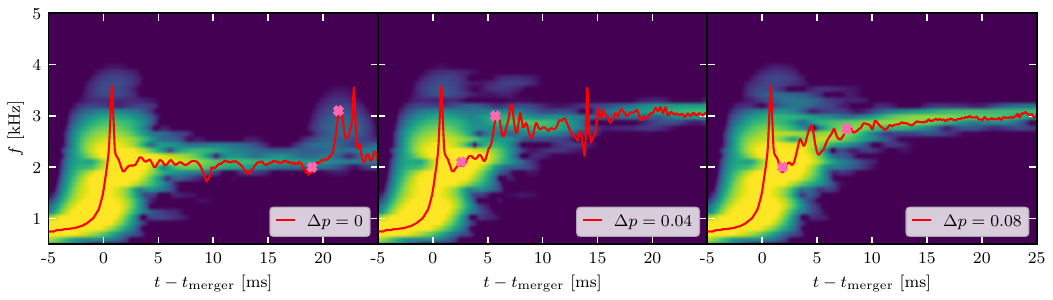}
    \end{minipage}
    \begin{minipage}{0.01\textwidth}
    \centering\vspace*{-0.5cm}\hspace*{-1.5cm}
        \includegraphics[scale=0.9]{prd_spectogram_scale.pdf}
    \end{minipage}
    \caption{Spectrogram of the GW signal from the low-mass merger. The red curve shows $f_\mathrm{GW}$ for that particular EoS, marked with time stamps from Fig.~\ref{fig:1.34_evolution}.}
    \label{fig:1.34_spectogram}
\end{figure*}

\begin{figure}
    \centering
    \includegraphics[scale=0.95]{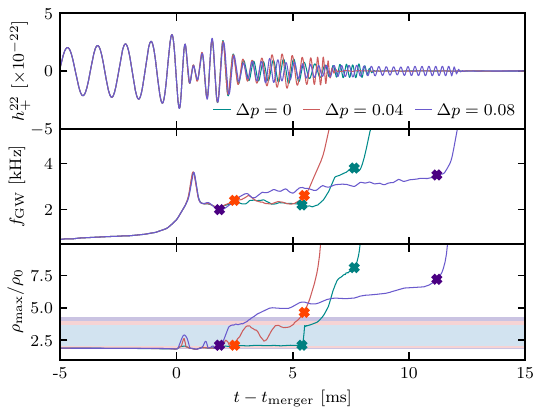}
    \caption{Evolution of the heavy mass merger. [Top] Extraction of $h_+^{22}$ strain at 100 Mpc distance from the merger. [Middle] Evolution of $f_\mathrm{GW}$ marked with the timestamps in alignment with time stamps marked in $\rho_\mathrm{max}$ evolution. [Bottom] Evolution of $\rho_\mathrm{max}$ marked with the start and end of the timespan of our interest. The shaded region hints at the mixed-phase region for the particular EoS.}
    \label{fig:1.38_evolution}
\end{figure}
\begin{figure}
    \centering
    \includegraphics[scale=1]{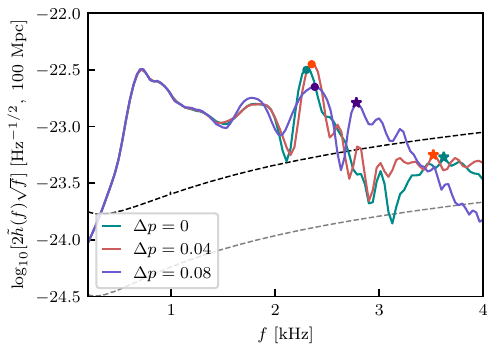}
    \caption{PSD of GW signals from heavy-mass merger at 100 Mpc. The $f_2^h$ and $f_2^q$ peaks are indicated with circle and star markers, respectively. The black and grey dashed lines are sensitivity curves of LIGO A+~\cite{Capote_2025} and ET-D~\cite{Hild_2011}.}
    \label{fig:1.38_psd}
\end{figure}
\begin{figure*}
    \centering
    \centering
    \begin{minipage}{0.95\textwidth}
    \centering\hspace*{-1.7cm}
        \includegraphics[scale=0.9]{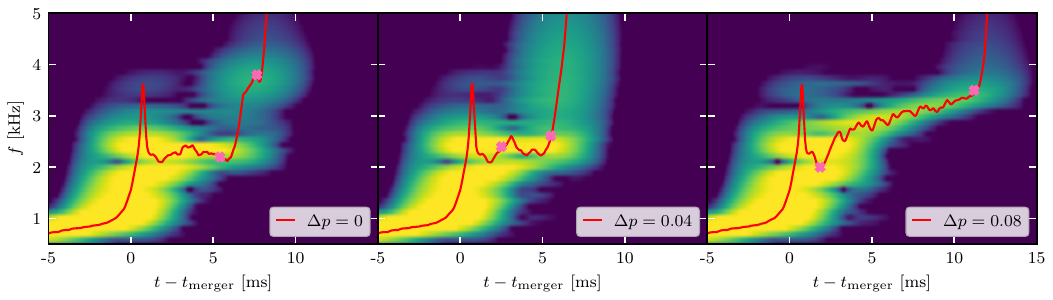}
    \end{minipage}
    \begin{minipage}{0.01\textwidth}
    \centering\vspace*{-0.5cm}\hspace*{-1.5cm}
        \includegraphics[scale=0.9]{prd_spectogram_scale.pdf}
    \end{minipage}
    \caption{Spectrogram of the GW signal from the heavy-mass merger. The red curve shows $f_\mathrm{GW}$ for that particular EoS, marked with time stamps from Fig.~\ref{fig:1.38_evolution}.}
    \label{fig:1.38_spectogram}
\end{figure*}

\begin{figure}
    \centering
    \includegraphics[scale=1]{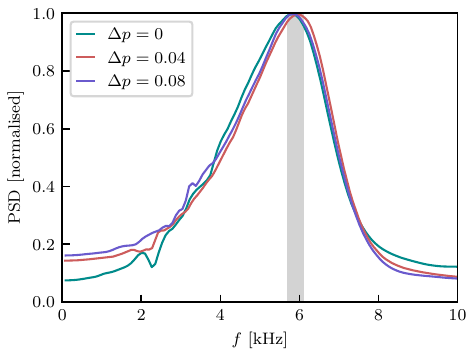}
    \caption{BH ringdown frequency estimates for the heavy-mass merger.}
    \label{fig:ringdown}
\end{figure}

\begin{figure}
    \centering
    \includegraphics[scale=0.92]{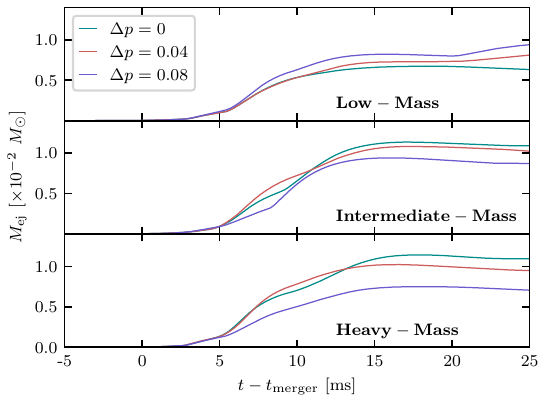}
    \caption{Evolution of $M_\mathrm{ej}$ from the event computed on the coordinate sphere at $r = 200~M_\odot (\simeq 295~\mathrm{km})$ from the merger.}
    \label{fig:ejecta}
\end{figure}

\begin{figure*}
\centering
\includegraphics[scale=1]{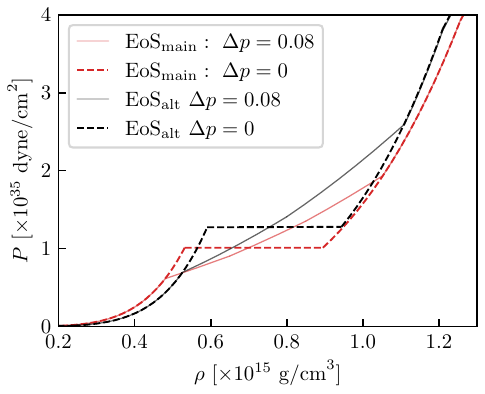}
\includegraphics[scale=1]{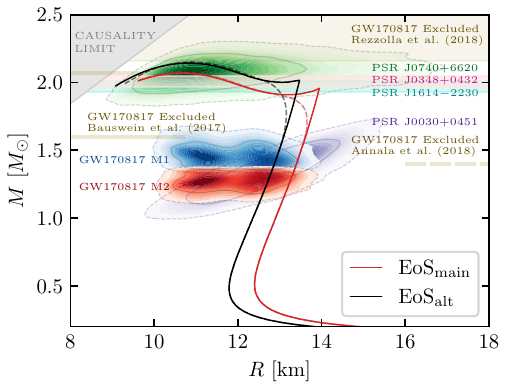}
\caption{ [Left] The pressure vs rest-mass density plot of the EoSs (alternate and main). [Right] The mass-radius ($M$--$R$) sequences corresponding to both EoS  satisfy the current astrophysical constraints~\cite{ozel_2010,Antoniodis_2013,GW170817,Abbot_2018,Abbot_2019,Miller_2019,Riley_2019,Miller_2021,Riley_2021,Annala_2018,Bauswein_2017_2,Rezzolla_2018_2,Lattimer_1990,Lattimer_2001}.}
\label{fig:eos_alt}
\end{figure*}

\begin{figure}
    \centering
    \includegraphics[scale=0.9]{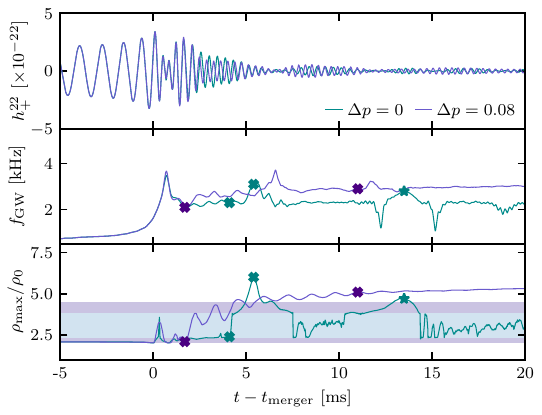}
    \caption{Evolution of the $2.84~M_\odot$ equal-mass merger for $\mathrm{EoS}_\mathrm{alt}$. [Top] Extraction of $h_+^{22}$ strain at 100 Mpc distance from the merger. [Middle] Evolution of $f_\mathrm{GW}$ marked with the timestamps in alignment with time stamps marked in $\rho_\mathrm{max}$ evolution. [Bottom] Evolution of $\rho_\mathrm{max}$ (in terms of $\rho_0$ = $2.51 \times 10^{14}~\mathrm{g/cm^3}$~\cite{glen}) cross-marked with the start and end of the timespan of our interest. The star marker is the indication of a second mini-collapse. The shaded region hints at the mixed-phase region for the particular EoS.}
    \label{fig:evolution_eos_alt}
\end{figure}
\begin{figure}
    \centering
    \includegraphics[scale=0.7]{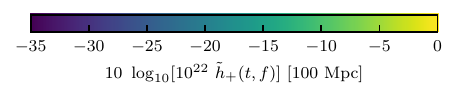}
    \includegraphics[scale=0.7]{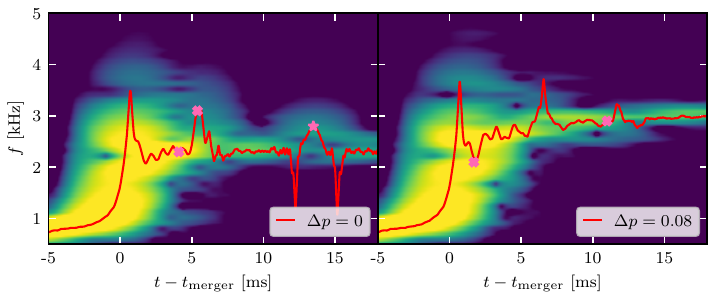}    
    \caption{Spectrogram of the GW signal. The red curve shows $f_\mathrm{GW}$ for that particular EoS, marked with time stamps from Fig.~\ref{fig:evolution_eos_alt}.}
    \label{fig:spectogram_eos_alt}
\end{figure}

\begin{figure}
\centering
\includegraphics[scale=1]{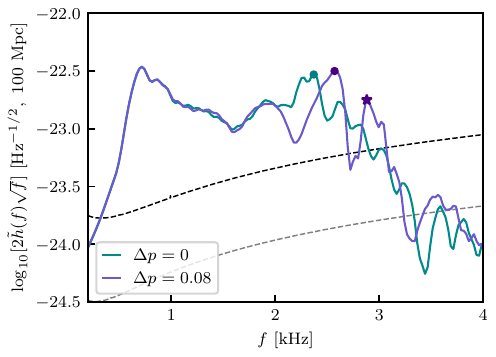}
\caption{PSD of GW signals from Fig.~\ref{fig:evolution_eos_alt} at 100 Mpc. The $f_2^h$ and $f_2^q$ peaks are indicated with circle and star markers, respectively. The black and grey dashed lines are sensitivity curves of LIGO A+~\cite{Capote_2025} and ET-D~\cite{Hild_2011}.}
\label{fig:psd_eos_alt}
\end{figure}

\begin{figure}
    \centering
    \includegraphics[scale=0.9]{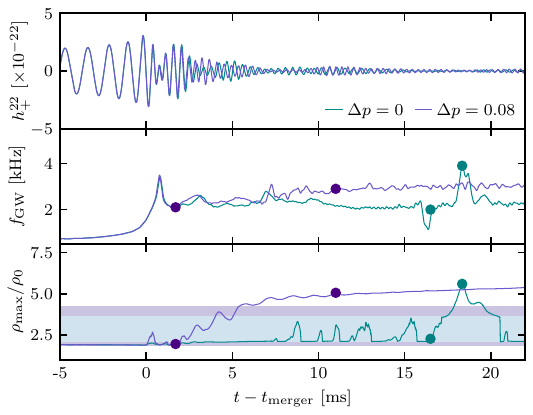}
    \caption{Evolution of the $2.72~M_\odot$ unequal-mass merger with $q=0.89$. [Top] Extraction of $h_+^{22}$ strain at 100 Mpc distance from the merger. [Middle] Evolution of $f_\mathrm{GW}$ marked with the timestamps in alignment with time stamps marked in $\rho_\mathrm{max}$ evolution. [Bottom] Evolution of $\rho_\mathrm{max}$ (in terms of $\rho_0$ = $2.51 \times 10^{14}~\mathrm{g/cm^3}$~\cite{glen}) cross-marked with the start and end of the timespan of our interest. The shaded region hints at the mixed-phase region for the particular EoS.}
    \label{fig:evolution_unequal}   
\end{figure}

\begin{figure}
\includegraphics[scale=0.7]{prd_spectogram_scale_2.pdf}
     \includegraphics[scale=0.7]{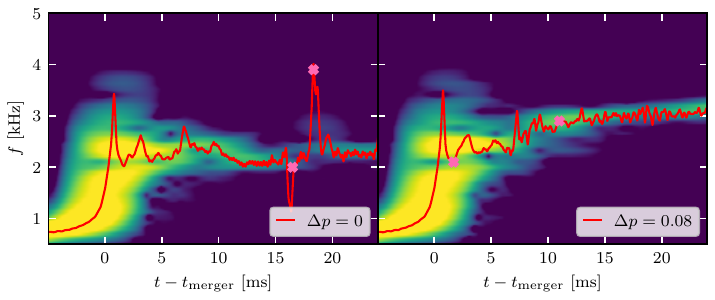}
     \caption{Spectrogram of the GW signal. The red curve shows $f_\mathrm{GW}$ for that particular EoS, marked with time stamps from Fig.~\ref{fig:evolution_unequal}.}
    \label{fig:spectogram_unequal}
\end{figure}

\begin{figure}
    \centering
    \includegraphics[scale=1]{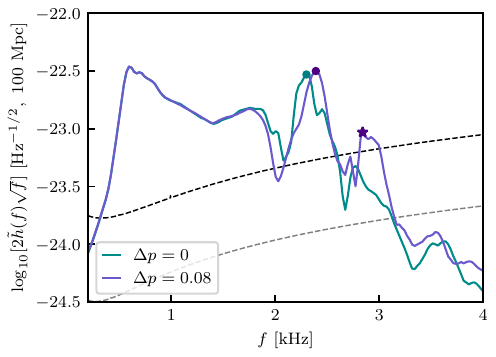}
    \caption{PSD of GW signals from Fig.~\ref{fig:evolution_unequal} at 100 Mpc. The $f_2^h$ and $f_2^q$ peaks are indicated with circle and star markers, respectively. The black and grey dashed lines are sensitivity curves of LIGO A+~\cite{Capote_2025} and ET-D~\cite{Hild_2011}.}
\label{fig:psd_unequal}
\end{figure}


\section{Analysis of Low Mass Merger}\label{1.34}

Figure~\ref{fig:1.34_evolution} monitors the evolution of the low equal mass mergers ($2.68~M_\odot$). Unlike intermediate-mass mergers, none of the remnants collapse to BH. For the EoS-$\Delta p_{0.08}$, we observe a gradual increase in $\rho_\mathrm{max}$, which leads to the introduction of pure QM at $\sim 6~\mathrm{ms}$. The two cross markers confirm that the steady increase in frequency results from HMNS converting to HMHS. After that, the core settles with hybrid matter, and the frequency saturates at a higher value. This signature is also reflected in the spectrogram plotted in Fig.~\ref{fig:1.34_spectogram} (right panel).

For the EoS-$\Delta p_{0.04}$, we observe a similar evolution as discussed above. However, the markers show that the rise in the $\rho_\mathrm{max}$ starts later and ends early with respect to EoS-$\Delta p_{0.08}$. This suggests a steeper rise in the density and, hence, a faster transition from HMNS to HMHS. This is also reflected in the frequency evolution and consequently in the spectrogram in Fig.~\ref{fig:1.34_spectogram} (middle panel). It is expected from the fact that lower $\Delta p$ in EoS has a softer mixed phase. Hence, the $\rho_\mathrm{max}$ will rise through mixed-phase faster and will have a quicker reach to the pure QM.

For EoS-$\Delta p_{0.0}$, the $\rho_\mathrm{max}$ evolution shows that the post-merger remnant endured HM for a prolonged period. The remnant evolves to HMHS for a short moment ($\sim3~\mathrm{ms}$) after $\sim20~\mathrm{ms}$ of evolution. The HMHS immediately experiences reverse PT and returns to the HMNS state. This results in a spike in the frequency evolution as well. In the spectrogram in Fig.~\ref{fig:1.34_spectogram} (left panel), it is reflected above the saturated value (associated with HMNS state) of post-merger frequency.

In Fig.~\ref{fig:1.34_psd}, we observe the PSD of the GW emissions from these systems at 100 Mpc. We found that $f_2^h$ frequencies, which is responsible for the remnant being in the HMNS state, increase with an increase in $\Delta p$. For the case of EoSs---$\Delta p_{0.08}$ and $\Delta p_{0.04}$, we observe two extra peak pairs indicated by star and diamond markers. We mark these peaks as $f_2^{q1}$ and $f_2^{q2}$. Since the remnant formed by EoS-$\Delta p_{0.0}$ remains in the HMNS state, such peaks are not expressed in the PSD. The expression of such peaks confirm the evidence of the hadron-quark PT, as discussed in the main text.

All the EoSs share identical values of $(K,\gamma)$ for the polytrope defining the hadronic part. However, the onset of PT in density increases with the decrease in $\Delta p$, after which the matter description changes. Hence, the mergers with lower $\Delta p$ experience a stiffer hadronic EoS before PT. In contrast, remnants with higher $\Delta p$ have already entered the mixed-phase part (or even the quark part). This results in the difference in post-merger dynamics (and hence different $f_2^h$ peaks) even for the phase where the remnant is in the HMNS state.

\section{Analysis of Heavy Mass Merger}\label{1.38}

Figure~\ref{fig:1.38_evolution} monitors the evolution of the heavy equal mass merger ($2.78~M_\odot$). Unlike intermediate-mass mergers, all the remnants collapse to BH. We observe the post-merger remnant from EoS-$\Delta p_{0.04}$ collapses the fastest among all three EoSs in comparison. The $\rho_\mathrm{max}$ in the remnant enters the mixed phase regime and oscillates there for a brief amount of time. Before reaching the pure QM, one of the bounces in the mixed phase regime triggers the remnant to enter exponential growth in $\rho_\mathrm{max}$, resulting in collapse into BH. Hence, the $\rho_\mathrm{max}$ spends a very short time in the pure quark regime. It can also be explained by the fact that the remnant had lost a lot of energy (and hence, angular momentum) through GW when $\rho_\mathrm{max}$ was evolving through mixed-phase. It can be observed that the post-merger GW shows the largest oscillation amplitudes in the time span $3$--$7~\mathrm{ms}$ with respect to other EoSs.

For the EoS-$\Delta p_{0.0}$, we observe $\rho_\mathrm{max}$ experiencing a sharp transition into the pure quark regime, resulting in the remnant entering into the HMHS phase. The remnant remains there for a very short amount of time, after which it enters exponential growth in $\rho_\mathrm{max}$, leading to collapse into BH. The $f_\mathrm{GW}$ also increases in this time span and halts at a higher frequency for a short period, this is also reflected in spectogram (Fig.~\ref{fig:1.38_spectogram}). However, in contrast to the low-mass and intermediate-mass mergers, the heavy-mass merger spectrogram shows a further shooting of the frequency, followed by a complete disappearance, indicating the remnant has experienced BH ringdown.

For the EoS-$\Delta p_{0.08}$, the $\rho_\mathrm{max}$ experiences a gradual growth until $\sim$$11~\mathrm{ms}$, after which, $\rho_\mathrm{max}$ enters the exponential growth part that indicates prompt collapse into BH. By looking at the collapse times, it is trivial to note that there is no relation between collapse times and $\Delta p$.

In the PSD, we observe that the $f_2^h$ frequency increases with an increase in $\Delta p$. However, it is difficult to resolve the difference as these remnants have short-lived HMNS phases, unlike low-mass and intermediate-mass mergers. For the EoS-$\Delta p_{0.08}$, the $f_2^q$ is expressed clearly. Subsequently, there are $f_2^q$ peaks from EoSs---$\Delta p_{0.0}$ and $\Delta p_{0.04}$, but are not strong enough. The reason for a weaker $f_2^q$ peak is due to the fact that both the remnants spent a shorter time in the HMHS phase.

In Fig.~\ref{fig:ringdown} we estimated the BH ringdown frequencies from the GW emission by considering the signal from the time stamp where the apparent horizon is detected in the system for the first time. Assuming that the BH ringdown is the main source for this part of the signal, the plot below indicates the frequency to be 6~kHz approximately. These frequencies are independent of $\Delta p$ in the EoS.

\section{Effect on ejecta mass}

In Fig.~\ref{fig:ejecta}, we report the ejecta mass ($M_\mathrm{ej}$) from the system. For intermediate-mass and heavy-mass mergers, lower values of $\Delta p$ in the EoS have led to higher $M_\mathrm{ej}$. However, for the low-mass mergers, we found the $M_\mathrm{ej}$ to be increasing with $\Delta p$. It has been shown that the post-merger remnant prompt collapsing into BH has a different dynamical evolution affecting the mass ejecta~\cite{Bauswein_2013_2,Lehner_2016,Radice_2018_2,Ciolfi_2017}, indicating that the end fate of the remnant adds additional complexity to the study. However, an in-depth investigation of the impact on kilonova afterglow and nucleosynthesis is possible.

\section{Consistency of GW Signature using a different EoS}

Using a different setup of EoS, we have shown the consistency of the two-fold GW signatures. In Fig.~\ref{fig:eos_alt} (left), we show that the different choices for hadronic and quark EoS were made to construct an alternate EoS ($\mathrm{EoS}_\mathrm{alt}$) with respect to the main EoS ($\mathrm{EoS}_\mathrm{main}$) used in the article. The newly constructed $\mathrm{EoS}_\mathrm{alt}$ supports larger $M_\mathrm{TOV}$, as shown in Fig.~\ref{fig:eos_alt} (right).

Figure~\ref{fig:evolution_eos_alt} shows the merger evolution of $2.84~M_\odot$ equal-mass merger. For $\mathrm{EoS}_\mathrm{alt}$-$\Delta p_{0.0}$, we observe the mini-collapse twice, as marked by cross markers and a star marker in the $\rho_\mathrm{max}$ plots. In both events, there is a local rise in $f_\mathrm{GW}$, which leaves an imprint on the spectrogram (Fig.~\ref{fig:spectogram_eos_alt} (left)) above the base post-merger frequency for HMNS. For $\mathrm{EoS}_\mathrm{alt}$-$\Delta p_{0.08}$, we observe a local peak in $f_\mathrm{GW}$ at $\sim 7~\mathrm{ms}$, but it is expressed due to destructive interference in GW strain, not due to PT.

In Fig.~\ref{fig:psd_eos_alt}, we plotted the PSD from the GW strains. For $\mathrm{EoS}_\mathrm{alt}$-$\Delta p_{0.08}$, we consistently find the $f^q_2$ peak (marked by star) beside the $f^h_2$ peak (marked by dot). For $\mathrm{EoS}_\mathrm{alt}$-$\Delta p_{0.0}$, we only observed the $f^h_2$ peak. There is no expression of the $f^q_2$ peak.

\section{Consistency of GW Signature using unequal mass merger}

We performed additional simulations of unequal mass merger ($M_\mathrm{tot}=2.72~M_\odot,q=0.89$) using the EoS set used in the main article, and we have shown the consistency of the two-fold GW signatures. For $M_\mathrm{tot}=2.72~M_\odot$, $q$ cannot be set to lower values ($\lesssim0.89$), because a quark seed will form in the high-mass companion initial data, which will create a pre-merger scenario we are not including in the study. We are only studying PT during or after the merger.

In Fig.~\ref{fig:evolution_unequal}, for $\mathrm{EoS}_\mathrm{alt}$-$\Delta p_{0.0}$, we consistently observe the mini-collapse in the post-merger, which leaves an imprint in $f_\mathrm{GW}$ and, on the spectrogram (above the base post-merger frequency for HMNS, as seen in Fig.~\ref{fig:spectogram_unequal} (left)). For $\mathrm{EoS}_\mathrm{alt}$-$\Delta p_{0.08}$, the $\rho_\mathrm{max}$ growth is smoother, leading to a smoother increase in $f_\mathrm{GW}$.

In Fig.~\ref{fig:psd_unequal}, we plotted the PSD from the GW strains. For $\mathrm{EoS}_\mathrm{alt}$-$\Delta p_{0.08}$, we consistently find the $f^q_2$ peak (marked by star) beside the $f^h_2$ peak (marked by dot). For $\mathrm{EoS}_\mathrm{alt}$-$\Delta p_{0.0}$, we only observed the $f^h_2$ peak, $f^q_2$ peak remained absent.

\bibliography{PRD}

\end{document}